# Expanding theory testing in general relativity: LIGO and parametrized theories

Lydia Patton

*Virginia Tech, USA*

### ABSTRACT

The multiple detections of gravitational waves by LIGO (the Laser Interferometer Gravitational-Wave Observatory), operated by Caltech and MIT, have been acclaimed as confirming Einstein's prediction, a century ago, that gravitational waves propagating as ripples in spacetime would be detected. Yunes and Pretorius (2009) investigate whether LIGO's template-based searches encode fundamental assumptions, especially the assumption that the background theory of general relativity is an accurate description of the phenomena detected in the search. They construct the parametrized post-Einsteinian (ppE) framework in response, which broadens those assumptions and allows for wider testing under more flexible assumptions. Their methods are consistent with work on confirmation and testing found in Carnap (1936), Hempel (1969), and Stein (1992, 1994), with the following principles in common: that confirmation is distinct from testing, and that, counterintuitively, revising a theory's formal basis can make it more broadly empirically testable. These views encourage a method according to which theories can be made abstract, to define families of general structures for the purpose of testing. With the development of the ppE framework and related approaches, multi-messenger astronomy is a catalyst for deep reasoning about the limits and potential of the theoretical framework of general relativity.

## 1. Theory dependence and theory testing

The recent exciting results in gravitational wave astronomy present novel challenges to philosophical accounts of theory testing.[1] The results promise to open new avenues of research. They will do so not just by racking up new detections, but by allowing researchers to formulate what Karl Popper called "interesting problems", problems that put the claims of a theory in jeopardy.[2] The results of the "detection" of gravitational waves are cited as "confirmation" of the predictions of general relativity. Suh discoveries are exciting, not just because of the results they confirm, but also because of the new doors for research and testing that they open. It is just as important to find ways to put a theory to the test as it is to confirm that theory. General relativity is not the only candidate theory of gravitation: scalar tensor theories, Brans-Dicke theory, massive graviton theory, Chern-Simons modified gravity, Einstein-Aether theory, MOND, TeVes, and DGP theory have all been proposed as alternatives to classical GR. It would be interesting, and scientifically productive, if the new events being detected could be cited as evidence for or against these theories.[3]

Still, any testing-based methodology comes with a challenge. The method of testing must be such that one can make logical, well founded inferences from tests of a theory to possible means of strengthening that theory. If our goal is to identify how a theory should be tested, and exactly which changes are warranted following testing, that will require identifying which of the hypotheses of that theory may be erroneous. It also requires finding a way to control testing of the theory: by showing which propositions of the background theory are independent of each other, for instance. That requires formal proofs of consistency and independence within axiomatized systems. The question of whether a

---

E-mail address: critique@vt.edu.

[1] See Thorne (1997) for an introduction to gravitational wave astronomy up to the early 1990s.

[2] Popper argues that "it is not the accumulation of observations which I have in mind when I speak of the growth of scientific knowledge, but the repeated overthrow of scientific theories and their replacement by better or more satisfactory ones" (Popper, 1989, p. 215). Critical examination of theories leads us to subject them to stress. We want to find the "interesting" problems that are hard cases for a theory - to put theories under harder and harder strain, until they break.

[3] Yunes and Pretorius 2009, Appendix A and Baker et al. 2017 are among recent papers drawing conclusions from the LIGO results about how to decide which among these theories are ruled out, and which is still viable.

model can be constructed that's independent of any given hypothesis of a theory is particulary crucial in gravitational wave astronomy. The methods of detection of gravitational waves require first constructing models of "candidate" waveforms, waveforms that are hypothesized to have the properties of the ones detected. These waveforms are used in computer simulations, which are in turn used for matched filtering of the signal. In practice, it does not seem possible to isolate any one hypothesis involved in gravitational wave modeling from any other. This is a problem, from the perspective of testing.

Philosophers of science will by now recognize another problem, to do with more well known forms of theory dependence in modeling. The fundamental claim at issue is whether the model being used is properly specified for the data under analysis. If a data model[4] assumes that the data modeled are independent and identically distributed (IID), but there is a trend in the data, regression analysis will not reveal the discrepancy, because the analysis itself has been formulated using a mis-specified model.[5] Similarly, if a dynamical model using differential equations assumes that the underlying phenomenon has a certain structure, it may or may not be possible to detect discrepancies from that assumption in testing. For instance, if a fluid model works on the assumption that the fluid is continuous, and the instruments used are not sensitive enough to detect or measure discontinuity, then those discrepancies (the actual granularity of the fluid) will not be detected in testing. This may or may not be significant, depending on the theory that is under investigation.

On the one hand, testing a theory requires finding tight logical connections between its abstract elements, so that researchers can identify weak hypotheses and claims, and can draw justified conclusions about how to revise a theory in the face of experimental findings in conflict with its claims. On the other hand, finding what Popper calls "interesting problems" that put the theory under strain may involve a more flexible approach. When searching for new ways to **test** a theory, it is crucial to consider, hypothetically, multiple ways the theory could be incorrect, and to identify the distinct ways the theory's claims could fail. That requires doing formal reasoning, not just about how to confirm the theory, but about how to show that it is wrong and exactly why.

The discussion below will demonstrate that the recent detections of gravitational waves can broaden the platform that gravitational wave astronomy provides for future theory testing and inference. The analysis below sketches one proposal for how this can be done, from Nicolás Yunes and Frans Pretorius, and provides a philosophical evaluation with a focus on theory testing and the growth of experimental knowledge. I will argue that a Carnapian tradition, represented by Howard Stein, Carl Hempel, and Carnap himself, provides a flexible account of theory testing and confirmation, and of the formal analysis of language and evidence, that makes it possible to widen a theory's empirical framework for testing.

1.1. Theory testing and confirmation

Logical relationships between elements of theoretical and observational vocabularies are fundamental to theory confirmation and testing. In order to assess whether a given experiment is a rigorous or 'severe' test of a theory, we might make certain counterfactual requirements of that test: that it control for type I and type II error, for instance.[6] Obtaining a clear definition of what it means to 'test' a theory is difficult, and it is even more difficult to distinguish confirmation (finding empirical instances that support a theory's claims) from testing (subjecting the theory to severe tests, i.e., to tests that the theory's hypotheses would fail if they were false)[7].

A most elegant analysis of the relation between theory, confirmation, and evidence is found in the work of Rudolf Carnap (Carnap 1936, 1988). Carnap made the problem appear clearly. But Carnap often is thought to have placed unduly stringent restrictions on the formulation of theoretical and observational vocabularies, and especially on the relations between the two.[8] Supposedly, the reduction of theoretical to observation sentences in (Carnap, 1936) requires first specifying universal relationships between theoretical and observational vocabularies. But this, as almost every major figure in the history of philosophy of science has observed, would be too strong (Stein, 1994; Hempel 1969, and so on). For example, as we will see in detail below, there is no general way to relate the equations of general relativity to the "observational vocabulary" of LIGO interferometry.

Thus, the focus increasingly comes to be on the question of how narrower and broader languages are related to one another to construct evidentiary relationships. In his early work, Carnap argued for relations between theoretical and observation languages, which allowed for the reduction of one to the other, and in turn for sentences in one to confirm sentences in the other. Hempel identifies a problem with this: it makes the languages of science seem as if they are entirely separate, when they are not. He observes.

> The assumption, in the standard construal, of an axiomatized uninterpreted calculus as a constituent of a theory seems to me, moreover, to obscure certain important characteristics shared by many scientific theories. For that assumption suggests that the basic principles of a theory – those corresponding to the calculus – are formulated exclusively by means of a 'new' theoretical vocabulary, whose terms would be replaced by variables or by dummy constants in the axiomatized calculus **C**. In this case, the conjunction of the postulates of **C** would be an expression of the type $\phi(\boldsymbol{t}_1, \boldsymbol{t}_2, ..., \boldsymbol{t_n})$, formed from the theoretical terms by means of logical symbols alone (Hempel 1969, p. 153).

Hempel observes that the theoretical terms are not really formed by stipulation in many cases. Rather, the previous theories inform the choice of variables, laws, concepts, structural elements, and explanations in the new theories. To Hempel, 'new' theories are not entirely new. The "theoretical scenario" in which theoretical terms and concepts are embedded is populated with terms that are interpreted using scientific knowledge and know-how accumulated before the new theory was conceived. Hempel calls this a 'pre-theoretical vocabulary'. When the new theory is stated, it is not a "totally uninterpreted system" other than the new theoretical terms and logical and mathematical symbols (Hempel 1969, p. 153). Instead, the new theoretical vocabulary is interpreted using the "pre-theoretical vocabulary" that defines, e.g., "masses, volumes, velocities, momenta, and kinetic energies" (Hempel 1969, p. 153).

---

[4] Here, in the sense of a model like the Normal/Gaussian distribution, the Bernoulli model, and so on. See Suppes 1966.

[5] Spanos and McGuirk have given a detailed analysis of "mis-specification testing," which is a set of tests designed to detect which of a family of *formal* models is appropriate to the data under analysis (Spanos & McGuirk, 2001).

[6] Deborah Mayo's work on statistical inference takes an approach focused on testing, but focuses on severe tests as criteria for rigorous learning through experimentation (Mayo, 2018). See (Spanos, 2011 and spa), and, of course, (Popper 1935, 1989).

[7] Here see especially Mayo, 2018 and Mayo, 1996.

[8] Despite well known criticisms of Carnap and of the logical empiricists on this score, there have always been those willing to see much of value in Carnap's approach. Richardson, 1998, Stein, 1994, Stein, 1992, Friedman, 1979 and even, to an extent, Glymour, 1975 and Glymour, 1980 can be read as in support of a broadly Carnappian framework. More recently, Lutz (Lutz 2012, 2017) provides stalwart defenses of the received view, backed up by some of the remarks in the introduction to Ott & Patton, 2018; an extensive yet sympathetic revision and reconstruction of the account was done by Dutilh Novaes and Reck 2017.

Hempel introduces a set of three vocabularies: theoretical, observational, and 'pre-theoretical'. His 'pre-theoretical' vocabulary is not limited to observation terms: terms and sentences describing wavelength, frequency, and so on are used to construct the 'internal principles' of a new theory developed about a related, but distinct subject matter. Like Imre Lakatos at around the same time, Hempel aims to "rationalize" "classical conventionalism" (Lakatos, 1970, p. 134). It is entirely possible within classical conventionalism to use previously gained knowledge as a *constraint on*, or even as a constitutive, rational basis for, the postulation of new internal principles of a theory.

Hempel's target is the position that "new" scientific theories begin from a blank slate of uninterpreted theoretical terms, logical and mathematical symbols, and fill in all the empirically meaningful 'slots' with novel theoretical concepts. It appears clear that Hempel's target was Carnap in particular.[9] In Carnap (1934), Carnap (1956), and other texts, Carnap provides a logical analysis of scientific language, resulting in a view that Friedman recently has called "structuralism without metaphysics" (Friedman (2011)).

> According to Carnap's account in the "Methodological Character" essay, only the observational terms of a scientific theory are semantically interpreted (by specifying observable properties and relations as their designata). The theoretical terms, by contrast, are semantically uninterpreted, and are only implicitly defined, in the sense of Hilbert, by the axioms and postulates of the relevant theory (e.g., Maxwell's equations for the electromagnetic field). Among these axioms and postulates, however, are mixed sentences or correspondence rules, which set up (lawlike) relationships among theoretical and observational terms; and, in this way, the theoretical terms and sentences receive a partial interpretation in terms of the connections they induce among observables. For example, Maxwell's equations, in the presence of suitable correspondence rules relating values of the electromagnetic field to actual measurements (of electric and magnetic intensities, and the like), generate observable predictions and thus have empirical content (Friedman (2011), pp. 253–4).

Along Hilbert's lines, and as late as 1956, Carnap takes the position that theories are frameworks or "scaffoldings" (Hilbert's term), which can given multiple interpretations. The theoretical terms, however, are not interpreted. Carnap writes, in response to Hempel's criticisms here and elsewhere,

> I agree with Hempel that the Ramsey-sentence does indeed refer to theoretical entities by the use of abstract variables. However, it should be noted that these entities are not unobservable physical objects like atoms, electrons, etc., but rather (at least in the form of the language which I have chosen in Carnap (1956)) purely logical-mathematical entities, e.g., natural numbers, classes of such, classes of classes, etc. Nevertheless it is obviously a factual sentence. It says that the observable events in the world are such that there are numbers, classes of such, etc., which are correlated with the events in a prescribed way and which have among themselves certain relations; and this assertion is clearly a factual statement about the world (Carnap (1963), p. 963).

Hempel argues, against this view, that there can be no re-statement of theoretical claims so that theoretical variables are entirely uninterpreted. In order to understand how a physical equation (say, Hooke's law or the ideal gas law) works, we need to have a 'pre-theoretical' understanding of terms like 'mass', 'volume', or 'pressure'. These terms encode, not only logical and mathematical symbols and relationships, but accumulated empirical knowledge. Hempel thus argues that theoretical vocabularies and concrete empirical interpretations of theories can't be disentangled as neatly as Carnap and Hilbert think they can.

Carnap himself (Carnap (1963)) was appreciative of Hempel's view on this matter. Similarly, I will concede to Hempel (for the sake of argument) that, when we are accounting for how a theory works to generate concrete explanations of phenomena, we cannot consider theoretical terms to be uninterpreted logical and mathematical symbols. In practice, it is necessary to use theories as systems of, not only symbols, but also accumulated empirical knowledge and know-how that constrains the application of those symbols.

If that previously gained knowledge is to inform future theory testing, however, there must be some logical framework within which we can locate the inferences drawn in theory testing and assessment. For instance, if we reason that all future physical theories in some domain must respect a measured physical parameter, that is, if the parameter must serve as a constraint on theories in that domain, we need a way to specify how the constraint will work. Which other measurements or quantities within the theory, for instance, will be affected? It is not possible simply to *state* that the parameter will serve as a constraint; one must *specify* how, in practice, the constraint will function, and what it will affect.

The earlier debate over Carnap's theoretical and observational languages turns, in this strain of work, into an analysis of how theoretical and observational vocabularies develop over time, and, in particular, how theoretical vocabularies and observational vocabularies are embedded within each other: not just observation within theory, but theory within theory, and observation within theory within theory. A previous (or rival) theory contains, not only an observational vocabulary, but internal principles or axioms of its own. Principles, laws, and definitions described the phenomena of the 'old' theory and constrained our analysis of them. From the viewpoint of a structural analysis, this means that we know what the phenomena *are* in the classical theory: we know what classical kinetic energy is, for instance, and what classical electron spin is. But when those concepts are imported into a 'new' theory - the quantum theory, for instance - they may undergo radical revision.[10] The incorporation of these terms of the 'pre-theoretical' - classical - language into the new quantum theory was like an experiment: how much of the 'old' theory can we incorporate without encountering anomalies and falsifications?

We can concede Hempel's point, but also recognize that, when we are considering the testing of a theory rather than its application, we are pushed to define theoretical terms increasingly *formally* in order to specify the abstract relationships that hold between terms, relations, and structures within the theory. This is true even if, in practice, the values of those terms must be constrained by evidence gained from empirical observation. Empirical evidence can constrain the range of value of a variable, for instance, and that variable can still be uninterpreted if it is specified conventionally.[11]

---

[9] See the related discussion of Hempel's "Theoretician's Dilemma" and Carnap's response, in Friedman (2011).

[10] See Salmon 1998, Büttner, Renn, & Schemmel, 2003, Patton, 2015, Kragh, 2012, and many more.

From a testing standpoint, there is potential in the Carnapian-Hempelian approach. As Popper emphasized, formal, logical relationships must be constructed in order to determine whether hypotheses are independent from one another, in order to identify the 'weak' premises of a model or theory. The Carnapian approach goes beyond this: If we look at theories as determining, not just classes of models as in the semantic view, but classes of languages including formal rules of inference, we can find flexible ways to control the testing of the theory, and to assess the empirical significance of a theory's claims (see Lutz, 2017 on empirical significance).

### 1.2. Stein on the theory-observation distinction

Howard Stein is well known for giving new life to the study of Isaac Newton. His work on Carnap is famous as well. Still, its potential remains untapped, and it is one of my aims for this paper to sketch a way that the new work on Carnap, much of which is indebted to Stein, can and should shape philosophy of science in the future. The account below will focus on "Some Reflections on the Structure of Our Knowledge in Physics" (Stein, 1994). Stein begins by revisiting Carnap's distinction between theoretical and observational languages.

> My own view is that in the rough sense Carnap was willing to adopt from the time he abandoned the more primitive versions of the empiricist thesis, there is no great difficulty in defining an "observational" vocabulary: an "observation-language" in Carnap's sense is the language in which we ordinarily conduct the business of daily life, and the only theory it is dependent upon is the theory that there are ordinary objects with such properties as we habitually ascribe to them. There are also systems of concepts of the sort that constitute the framework of fundamental physical theories; so, referring again to my example, I may say that a book like *Raum – Zeit – Materie* [by Hermann Weyl] demonstrates the existence of theoretical vocabularies distinct from the observational. Thus I argue, on the basis of these crude and banal considerations, that Carnap was right to make and to emphasize this distinction. I also believe that his philosophic career consists to a considerable degree in a series of genuinely instructive attempts to do better justice to the character of the distinction (Stein, 1994, p. 638).

Stein goes on to identify a specific difficulty:

> But I think too that there was a fundamental bar to success along any of the routes Carnap essayed. For he always assumed that "the observation language" is more restricted than, and included in, a total language that includes an observational part and a theoretical part, connected by deductive logical relations. And this, I think — I do not say by virtue of some basic principle I can identify, but simply, at the present time, de facto — is not the case: there is no department of fundamental physics in which it is possible, in the strict sense, to deduce observations, or observable facts, from data and theory. So I suggest that the principal difficulty is not that of how to leave the theory outside the laboratory door, but that of how to get the laboratory inside the theory (Stein, 1994, p. 638).

Stein argues that the fundamental problem is that there is no straightforward relation between two vocabularies, observational and theoretical, that allows for the reduction of one to the other, or the deduction of one from the other.

Stein makes the following claims:

1. In many cases, we do not know how to deduce observations from theory.
2. There, we cannot employ Carnap's distinction between theoretical and observational vocabularies to "reduce" one to the other in order to construct a confirmation relation.
3. Moreover, we cannot construct a language in which a more restricted observational vocabulary is contained within a universal theoretical vocabulary, which will capture all the inferential relations necessary to theory confirmation and testing.
4. Thus, it is not that Carnap was wrong to make the distinction between theoretical and observational vocabularies, but that he was not able to show that that distinction was supported by an embedding of a restricted observation language within a universal theoretical language.

Here Stein makes a surprising move. He argues that demonstrating the links between observational and theoretical vocabularies requires increased clarity at the level of the *theoretical*. "Getting the laboratory inside the theory," as he puts it, requires going back to Carnap's "first volume": the axiomatic or law-governed first principles of a theory. Stein uses Newtonian classical mechanics as an example: Newtonian rational mechanics still is a progressive research program, even as Newtonian classical mechanics has been recognized to be less accurate than relativity and the quantum theory in certain regimes. We need not accept a strict distinction between theoretical and observational vocabularies to accept that progress in "purely mathematical theory" may go on even when removed from its physical application - or to accept that that purely mathematical theory may in turn be embedded in another empirical theory later, or in a plurality of empirical theories (Stein, 1994, p. 650).

Erik Curiel (Curiel, 2020) identifies a framework for showing how theoretical and experimental data relate to one another, explicitly inspired by Stein's urging to get "the laboratory inside the theory". The common thread of these approaches is the refusal to collapse the languages that Hempel so carefully teased apart: the theoretical, the observational, and the pre-theoretical. Curiel makes several distinctions that will be crucial in what follows.

> The template in a framework for equations of motion and other mathematical relations I call *abstract*. Canonical examples are Newton's Second Law, the Schrödinger equation in non-relativistic quantum mechanics, and so on. Structure and entities at the highest level of a theory formulated in a given framework I will call *generic*. In particular, generic structure has no definite values for those quantities that appear as constants in the theory's equations of motion and other mathematical relations. The symbol '$k$' appearing in the generic equation of motion of an elastic spring modeled as a simple harmonic oscillator, $\ddot{x} = -\frac{km}{x}$, denotes Hooke's constant (the coefficient of proportionality of a force applied to the spring and the resulting displacement from its equilibrium position), but possesses no fixed value, and the same for the mass $m$ ... One obtains *specific* structure by fixing the values of all such constants in generic structure, say $m = 1$ and $k = 5$ (in some system of units) for the elastic spring. This defines a *species* of physical system of that genus, all springs with those values for mass and Hooke's constant. One now has a determinate space of states for systems of that species, and a determinate family of dynamically possible evolutions, viz., the solutions to the specific equations of motion, represented by a distinguished family of paths on the space of states (Curiel, 2020, §7).

---

[11] If we say "Let height h range between 2 and 4 feet", for instance, h is not "interpreted" in the semantic sense just because it has empirical 'content'.

Once the family of possible physical systems has been identified, one can begin to look for family members. These will come in the form of "a *concrete* model": "a collection of experimentally or observationally gathered results structured and interpreted in such a way as to allow identification with an individual model" (Curiel, 2020, §7). The significant point here is that the identification between concrete and individual models "embodies the substantive contact between theory and experiment required to comprehend the full epistemic content of a theory" (Curiel, 2020). But this identification itself has nothing to do with experiment. It is an identification between elements of two distinct *languages*. That is emphatically not to say that these two languages must be explicitly set up before any meaningful empirical content can be found. But, if we are to understand the empirical content or significance of a theory fully, we must know whether the terms, concepts, propositions, and models we are using are abstract, generic, concrete, or individual, and we must have the tools needed to make justified comparisons between elements of these distinct classes.

The parametrized versions of the theory of general relativity that we will discuss below are explicitly intended to identify which parts of the theory are abstract, which generic, which concrete, and how they work together in elaborating and testing the theory. The example of LIGO is an excellent way to cash out what Stein and Curiel mean by getting "the laboratory inside the theory".

## 2. The LIGO results as tests of general relativity

The detection of gravitational waves by the LIGO (Laser Interferometer Gravitational-Wave Observatory), operated by Caltech and MIT, has been acclaimed as confirming Einstein's prediction a century ago that gravitational waves propagating as ripples in spacetime would be detected. The detection is cited as evidence in favor of general relativity. This paper will not question the detection itself, but rather, will investigate how the LIGO data and methods can best be used as a platform for future scientific investigation and inference. The method will be to analyze whether the framework proposed for analysis of the theory demonstrates that the theory of general relativity (GR) was tested by the data and inferences from it, which is distinct from the claim that the observations can be shown to confirm the theory.

The paper builds on two strands of research, one scientific and one philosophical. The scientific resource is recent work by Nicolás Yunes and Frans Pretorius, on a certain framework of assumptions that they see as encoded in the current methods of gravitational wave astronomy (Yunes & Pretorius, 2009). In a later paper, Yunes, Kent Yagi, and Pretorius refer to Karl Popper as the motivation for their approach:

The social scientist and epistemologist Popper argued that scientists can never truly "prove" that a theory is correct, but rather all we can do is disprove, or more accurately constrain, alternative hypotheses. The theory that remains and cannot be disproven by observations becomes the status quo. Indeed, this was the case for Newtonian gravity before the 1900s, and it is the case today for Einstein's theory of general relativity (GR). The latter has been subjected to a battery of tests through Solar System, binary pulsar and cosmological observations, with no signs of failure. These tests, however, cannot effectively probe the *extreme gravity regime*: where the gravitational field is strong and dynamical, where the curvature of spacetime is large, and where characteristic velocities are comparable to the speed of light. The [LIGO results] allow for just that (Yunes, Yagi, & Pretorius, 2016, pp. 084002−1 and 084002−2).

In what follows, I will argue that Yunes & Pretorius, 2009 and Yunes et al., 2016, in providing a scientific analysis of the LIGO results with respect to theory testing, go well beyond Popperian falsification. They provide a method for showing how a theory can become the node of a broader network for heuristic testing. They build a theory within the theory of GR, a parametrized set of models that allow for more rigorous testing of hypotheses about *deviations* from the theory's predictions and structure.

The philosophical discussion focuses on Yunes's and Pretorius's construction of a family of related theories, generated by modeling consequences of, and deviations from, fundamental assumptions. The paper traces a tradition, with beginnings in the work of Rudolf Carnap and Carl Hempel, of considering scientific formal reasoning as constructing a network of embedded formal languages. These languages reflect observational vocabularies, methods of testing, purely mathematical reasoning, and even conventional specification of laws of nature. Analysis of formal languages shows how they function, in practice, to allow for narrower or broader analysis of data: e.g., by confirming or falsifying a background theory, or by constraining the values of variables and the reach of laws. When used as a framework for testing, specifying the formal vocabulary of theories and defining families of theories under consideration allows for rival theories to be compared and for some to be discarded.

In keeping with this special issue on the work of Howard Stein, I focus on Stein's analysis of the role of axiomatic, formal reasoning in theory development and testing. Stein advised us to "get the laboratory inside the theory" (Stein, 1994, p. 638; Curiel, 2020). Moreover, he, like Hempel, thought that observational and theoretical vocabularies were preserved within successive theories. I take inspiration from his analysis of internal theoretical relationships as being the basis for showing how a theory can be tested more rigorously, and can be compared to rivals.

The conclusions I will draw are simple: that internal, formal reasoning paradoxically can broaden a theory's empirical reach, and that it can do so by removing barriers to broader empirical testing that have been encoded into a theory's formal structure. Barriers can include relations constructed to aid in confirmation, like constraints on parameter values, methods for modeling and simulation, and linkages between measurement and inference, that narrow a theory's scope for testing in practice. The sections above have evaluated one philosophical approach to these questions, found in Carnap, Hempel, and Stein. The sections following will focus on the recent LIGO results, and on the analysis of them by Yunes, Yagi, and Pretorius. I will argue that Pretorius, Yagi, and Yunes provide an analysis very well suited to the Carnap-Stein framework for testing.

### 2.1. Template-based searches in gravitational astronomy: modeling physical parameters

Multi-messenger astronomy is the search for signals from astronomical events by using multiple sites of detection, searching for signals from a target system and some other source: neutrinos, for instance.[12] Using multiple sources allows for triangulation of the signal, and, in other ways, improves inferences from data. A most striking feature of multi-messenger astronomy, when working on gravitational waves in particular, is the fact that quite a bit of mathematical structure is necessary to confirm a detection of a compact binary coalescence (CBC): in this case, a pair of merging black holes.

---

[12] I am grateful to an anonymous reviewer for pressing clarification on this point.

The data emerging from advanced LIGO (Laser Interferometer Gravitational-Wave Observatory) techniques are differences in the lengths of an interferometer's 'arms', which in the case of the aLIGO (advanced LIGO) projects currently running in Hanford, Washington, and Livingston, Louisiana, are 2 km long laser paths in a vacuum. (Advanced LIGO or aLIGO is so called to distinguish it from "Initial Ligo" or iLIGO, involving an earlier, less sensitive generation of detectors.) Based on hypotheses about the polarization of gravitational waves hitting the interferometer, physicists infer the plus and cross polarizations of the waveforms. Template banks are generated using estimates of the physical parameters of the black holes, including their mass and spin, the chirp mass, as well as parameters of the compact binary coalescence (CBC), including the frequency and period of the black holes' orbit around each other (orbital frequency and orbital period).

There are three stages to a black hole merger: inspiral, in which the black holes come within each others' gravitational fields and begin to spin around each other more and more quickly; merger, in which the black holes coalesce; and ringdown, in which the black holes have merged and the dynamics are now approximately those of one body.[13] Einstein's field equations for general relativity cannot be solved analytically in the inspiral and merger stages. Nonetheless,

> The inspiral phase can be modeled very well by the restricted PN approximation… In this approximation, the amplitude of the time-series is assumed to vary slowly relative to the phase, allowing its Fourier transform to be calculated via the stationary-phase approximation (Yunes & Pretorius, 2009, pp. 122003–6).

The "restricted PN approximation" is the restricted post-Newtonian approximation, "an umbrella term for updating Newton's equations perturbatively using a variety of series expansions, most notably with respect to the dimensionless source velocity $\frac{v}{c}$" (Holst, Sarbach, Tiglio, & Vallisneri, 2016, p. 520).

The merger phase is much more difficult, as it "cannot be modeled analytically by any controlled perturbation scheme" (Yunes & Pretorius, 2009, pp. 122003–7). The key here is that the merger phase not only cannot be derived as a direct solution to the field equations of general relativity, it cannot be modeled as a perturbation of known waveforms, for instance, damped sinusoids. Yunes and Pretorius, the latter of whom made significant earlier breakthroughs in the ability to provide numerical relativity solutions to the field equations (Holst et al., 2016, pp. 524, 542), suggest modeling the merger phase as an interpolation between the inspiral and ringdown (Yunes & Pretorius, 2009, pp. 122003–7).

The field equations of general relativity cannot be solved directly in the dynamical, strong-field region of gravity. More than that, observations in the dynamical strong-field region have been lacking to date (Yunes & Pretorius, 2009, pp. 122003–2). But there has been reason to believe that such observations were possible since 1974, when

> Russell Hulse and Joseph Taylor Jr. discovered the binary pulsar PSR B1913 + 16, whose orbit was later shown by Taylor and Joel Weisberg to shrink in remarkable agreement with the emission of gravitational radiation, as predicted by Einstein's quadrupole formula. This discovery, which led to a Nobel Prize in Physics in 1993 for Hulse and Taylor, was the first clear if indirect demonstration of the existence of [gravitational waves]. In fact, the precise timing analysis of PSR B1913 + 16 (and of other similar pulsars) reflects and demonstrates a broader range of general-relativistic corrections, which are computed in the $post-Newtonian\ approximation$ (Holst et al., 2016, p. 520).

Bolstered by the success of Hulse and Taylor, early researchers in gravitational wave astronomy, including Pretorius, Rainer Weiss, Kip Thorne, Robert Wald, Yvonne Choquet-Bruhat, and others, undertook to show how interferometer data could be shown to constitute a detection of waveforms emanating from merging black holes.

> As the BHs get closer to each other and their velocities increase, the accuracy of the PN expansion degrades, and eventually the full solution of Einstein's equations is needed to accurately describe the binary evolution. This is accomplished using numerical relativity (NR) which, after the initial breakthrough, has been improved continuously to achieve the sophistication of modeling needed for our purposes. The details of the merger and ringdown are primarily governed by the mass and spin of the final BH [black hole]. In particular, the final mass and spin determine the (constant) frequency and decay time of the BH's ringdown to its final state. The late stage of the coalescence allows us to measure the total mass which, combined with the measurement of the chirp mass and mass-ratio from the early inspiral, yields estimates of the individual component masses for the binary (Abbott et al. (2016)).

The quadrupole formula, the chirp mass, and the non-constant orbital period of the black hole mergers under analysis cemented the use of gravitational wave observations as confirmation of general relativity. The "chirp mass" is modeled by an equation modeling the falloff in mass as massenergy radiates from a compact binary coalescence. The "orbital period" of a binary black hole system is the time it takes one black hole to complete an orbit about the other. If Newton's theory were correct, the orbital period would be constant. A waveform from which one can estimate, and then confirm, a non-constant orbital period rules out Newtonian gravity, and is consistent with general relativity.

Those studying black hole dynamics have developed multiple methods of confirming detections of black holes using interferometer data. Simulated waveforms using estimates of the system parameters are used to construct a signal space covered by a minimal set of candidate waveforms, after which the candidate waveforms are used for matched filtering.[14] It took decades of labor for the process described in the last sentence to be made possible. The mathematical progress made, summarized in Holst et al., 2016, has allowed for the merger phase to be modeled using a combination of analytical methods (solutions of the Einstein-Hilbert equations) and numerical relativity or 3 + 1 methods (see Abbott et al., 2016).

---

[13] "We divide the coalescence into three stages: (i) the inspiral; (ii) the plunge and merger; (iii) the ringdown. In the first stage, the objects start widely separated and slowly spiral in via GW radiation-reaction. In the second stage, the objects rapidly plunge and merge, roughly when the object's separation is somewhere around the location of the light ring. In the third stage, after a common apparent horizon has formed, the remnant rings down and settles to an equilibrium configuration. Note that this classification is somewhat *fuzzy*, as studies of numerical simulation results have shown that a sharp transition does not exist between them in GR" (Yunes & Pretorius, 2009, pp. 122003–6).

[14] See, e.g., Capano et al., 2016. My thanks to Zachary Shifrel for pointing out this paper.

Numerical relativity is a technique that allows for the simulation of a waveform: one that *would* emanate from a black hole coalescence with specific system parameters. Some of the parameters turn out to be more fundamental to the generation of the simulated waveform than others. For instance, the mass and spin of the final single black hole largely determine the and decay time from the merger to the ringdown phase.

The chirp mass of the binary drives the simulation of the inspiral phase. As Yunes and Pretorius note, the merger can be treated as an 'interpolation' between the inspiral and ringdown. The ringdown phase is modeled using 'quasi-normal modes' or damped sinusoids, which are understood from observations of the radiation of neutron stars.[15] But the inspiral is a different story: "the phase evolution [of the inspiral phase] is driven by a particular combination of the two masses, commonly called the chirp mass" (Abbott et al. (2016), pp. 241102–3). The chirp mass equation describes the diminution of the mass of the system over time, as massenergy is radiated away as gravitational waves:[16]

$$\mathcal{M} = \frac{(m_1 m_2)^{\frac{3}{5}}}{M^{\frac{1}{5}}}$$

In this equation, $m_1$ and $m_2$ are the masses of the black holes, M is the total mass, and $\mathcal{M}$ is the chirp mass. In time, the chirp mass is estimated by the following equation:

$$\mathcal{M} \simeq \frac{c^3}{G} \left[ \frac{5}{96} \pi^{-\frac{8}{3}} f^{-\frac{11}{3}} \dot{f} \right]^{\frac{3}{5}}$$

Here, f is the gravitational wave frequency and $\dot{f}$ is its derivative with respect to time (Abbott et al., 2016, pp. 241102–3).

The template banks for LIGO require specifying an estimated chirp mass ahead of time. It is one of the fundamental parameters of binary black hole systems, and one of the subset of the physical parameters that drive the final waveform (Abbott et al. (2016), pp. 241102–3). The chirp mass equation is solved in part by estimating the mass of the black holes and the frequency of the waves hitting the detector.

Given the aLIGO methodology, it is crucial to distinguish two sets of parameters. The first are the *hypothetical physical parameters* that are used to construct a simulated numerical relativity (NR) waveform. The second are the *estimated physical parameters* that are inferred from the waveform that is selected as the best confirmed from the template bank: that is, the detected waveform. The LIGO collaboration does not use these terms, but they are useful to distinguish them from each other in characterizing the methods used.

Abbott et al. (2016) focuses on estimating physical parameters of the first aLIGO detection, GW150914. These estimates can be given a complex set of constraints, including: theoretical constraints (from GR, Hamiltonian mechanics, and Fourier analysis, for instance); physical constraints (e.g., from knowledge about neutron star binaries); and empirical limits derived from the properties of the detected waveform. The latter constraints are determined by Bayesian analysis of the confidence interval within which the detection can be confirmed.

The hypothetical physical parameters used to derive candidate waveforms from the template bank are used in physical reasoning: with GR, and with the post-Newtonian approximation and the quadrupole formula. That physical reasoning is used to simulate binary black hole systems and to construct hypothetical waveforms that might be detected as emanating from systems with specific physical parameters. When a single waveform is selected for confirmation from the template bank based on the data, the likelihood ratio is used to confirm that waveform as evidence for the detection of a system with those specific physical parameters.

Estimating the physical parameters of a system is done in post-data analysis, using admirably open source, customized methods for the LIGO collaboration (Abbott et al., 2016, e.g.). There are ongoing projects investigating the possible effects of "systematic" errors on the resulting analysis, whether theoretical or statistical.[17] These projects are conceived as part of the collective work of the LIGO collaboration. These projects do not focus on the possible structural assumptions encoded in the models that have to do with the dynamics of GR, or with assumptions about the time-dependent variation of the chirp mass, for instance.

The proccess of finding candidate waveforms to confirm with respect to the data is freighted with theoretical assumptions - by design. Moreover, the assumptions about the time-dependent dynamics of the system are co-variant. The "chirp mass", a time-dependent massenergy function, is a case in point. Building a candidate (hypothetical) waveform requires providing varying hypothetical physical values for the variables involved. Once a candidate waveform is identified as the most likely in post-data analysis, the chirp mass of the system is now considered a *physical* parameter of that system. This allows researchers to treat the chirp mass as a "parameter" that can be estimated using the interferometer data. But the chirp mass was not measured directly in that initial process. It is used as one of a number of hypothetical physical parameters in constructing a set of candidate waveforms, and then it is confirmed, along with all the other hypothetical parameters, when the candidate waveform is judged the most likely.

The chirp mass function is co-variant with physical variables like the initial masses of the two black holes (or other astronomical bodies) and with the final, single mass post-merger, in the ringdown phase. And the chirp mass function has a very significant, almost dominant effect on the amplitudes of the waveforms (see, e.g, Thorne, 1997, p. 14). This has led to the perception that one can "read off" the chirp mass from the detected waveform. Of course, to read off these features from the waveform, one has to get to the accurate waveform first!

Getting to the accurate waveforms is done, now, in a process that includes matched filtering that depends on a template bank of candidate waveforms, all of which are produced using hypothetical assumptions about the chirp mass and other parameters and functions, and assumptions about the dynamics of the system of which these elements are descriptions. If the aim is to *test* general relativity, in the context of rival theories and of other theories as yet unconceived, how should researchers treat these assumptions? In this section, we have focused on how substantive, hypothetical physical parameters are embedded in the modeling and simulation methods of aLIGO. In the section following, we will see how how assumptions about GR *itself* are "baked in" to the templates and inferences in model-based template searches. For a more robust testing framework for aLIGO to be built requires analysis of how these assumptions can be made 'generic', in Curiel's sense: for instance, so that they define a class of structures that can be tested to determine which of the candidate set is confirmed by the observations.

---

[15] "Practically every stellar object oscillates radially or nonradially, and although there is great difficulty in observing such oscillations there are already results for various types of stars (O, B, …). All these types of pulsations of normal main sequence stars can be studied via Newtonian theory and they are of no importance for the forthcoming era of gravitational wave astronomy. The gravitational waves emitted by these stars are extremely weak and have very low frequencies … This is not the case when we consider very compact stellar objects i.e. neutron stars and black holes. Their oscillations, produced mainly during the formation phase, can be strong enough to be detected by the gravitational wave detectors (LIGO, VIRGO, GEO600, SPHERE) which are under construction. In the framework of general relativity (GR) quasi-normal modes (QNM) arise, as perturbations (electromagnetic or gravitational) of stellar or black hole spacetimes. Due to the emission of gravitational waves there are no normal mode oscillations but instead the frequencies become 'quasi-normal' (complex), with the real part representing the actual frequency of the oscillation and the imaginary part representing the damping" (Kokkotas and Schmidt 1999; Introduction).

[16] See equation (3) of Abbott et al., 2016, p. 241102–3, for a more detailed version.

[17] Abbott et al. (2016), Ghosh, Del Pozzo, and Ajith (2016), and others. I am grateful to Heta Patel for informing me of Ghosh's work.

## 2.2. The parametrized post-Einsteinian framework

Frans Pretorius and Yvonne Choquet-Brouhat were among those whose work allowed for the links between numerical relativity methods and solutions to the linearized field equations to be constructed. More recently, a paper by Yunes and Pretorius (2009) argues that the very methods that Pretorius participated in constructing encode what Yunes and Pretorius call "fundamental theoretical bias", which they contrast with "modeling bias".[18] A modeling bias might include the choice to model a fluid as continuous rather than granular, for instance, or to make the usual simplifying assumptions. In the case of the dynamics of astronomical binary systems (e.g. orbiting black holes or neutron stars), a *modeling bias* is a case "where the preconception relates to physical assumptions to simplify the solutions considered (e.g. that all binaries have circularized prior to merger), or unverified assumptions about the accuracy of the solution used to model the given event" (Yunes & Pretorius, 2009, pp. 122003–1). Modeling bias arises in any number of situations in physics and is well known.[19]

The "fundamental bias" analyzed by Yunes and Pretorius is more specific: in the case of general relativity, it is the assumption that the Einstein field equations are valid in the strong-field regime (Yunes & Pretorius, 2009, pp. 122003–1). They note:

> Systematic errors created by fundamental bias may be as large as, if not larger than, those induced by modeling bias, as waveforms could deviate from the GR [general relativity] prediction dramatically in the dynamical strong field, if GR does not adequately describe the system in that region. This is particularly worrisome for template-based searches, as the event that will be ascribed to a detection will be the member of the template bank with the largest SNR [signal to noise ratio]. Given that GR is quite well tested in certain regimes, many sources cannot have deviations so far from GR as to prevent *detection* with GR templates (albeit with lower SNR). Thus, if templates are used based solely on GR models, although the corresponding events may be "heard," any unexpected information the signals may contain about the nature of gravity will be filtered out (Yunes and Pretorius (2009), pp. 122003–1).

A template-based search uses matched filtering to select the best candidate waveform from a signal space that's been covered ahead of time with a minimal set of possible waveforms generated using a mixture of numerical relativity, analytical methods, and estimated values for the parameters of the source. The worrying possibility is that an event with a signal to noise ratio that's picked up by the template based search will be close to the predictions of general relativity, but the ways in which it deviates from GR will be masked by the templates themselves.

Two questions motivate Yunes's and Pretorius's approach:

1. Suppose gravity is described by a theory differing from GR in the dynamical, strong-field regime, but one observes a population of merger events filtered through a GR template bank. What kinds of systematic errors and incorrect conclusions might be drawn about the nature of the compact object population due to this fundamental bias?
2. Given a set of observations of merger events obtained with a GR template bank, can one quantify or constrain the level of consistency of these observations with GR as the underlying theory describing these events? (Yunes and Pretorius (2009), pp. 122003–3).

We might wish to see what Yunes and Pretorius call a "fundamental theoretical bias" as a framework of assumptions, applied to modeling.[20] If the entire modeling framework you have constructed is not suitable for testing a theory, but only for confirming it, then you are not putting the claims of the background theory in any jeopardy. You may then find that the observations you've made are consistent with the theory, but have not controlled for certain types of error. The question is of the epistemic foundation for accepting or rejecting the *model* that you have used. Do you have justification for the claim that the hypotheses of the model are properly independent of the background assumptions of the theory under testing, or not?

Testing requires constructing a framework under which we can show whether the hypotheses of a model are independent of, or dependent on, the theory under test. The framework I advocate allows for the construction of a more robust framework for testing through *formal* reasoning. We can identify distinct observational, theoretical, and "pre-theoretical" languages within a theory, all of which can be analyzed as logically dependent on or independent of the others. In particular, we might be able to construct formal languages within a theory specifically for the purposes of testing, and for the purpose of showing which of the assumptions of a theory are independent of the others and can therefore be a strong basis for empirical testing.

Yunes and Pretorius propose a "parametrized post-Einsteinian (ppE) framework" along these lines.[21] LIGO astronomy is based on the assumption that "a compact object such as a BH [black hole] or NS [neutron star] spiraling into a supermassive BH will emit waves that carry a map of the gravitational field of the central BH" (Yunes and Pretorius (2009), pp. 122003–2). That map is partly drawn by investigating the polarization of the signal, which is similar to investigating the E and B fields by determining the polarization of light. Yunes and Pretorius focus on the 'plus' and 'cross' polarizations of gravitational waves.

---

[18] That fact should alert us that the term 'bias' is being used here to mean 'deviation resulting from an assumption or structural feature', which is not necessarily negative.

[19] The use of hypothetical physical parameters in template-based modeling could be seen as a kind of modeling bias, or not, depending on how you see the methods and conclusions unfolding in practice.

[20] I am grateful to an anonymous reviewer for prompting this reflection.

[21] Chapter 3 of (Will, 2014) provides a detailed examination of GR as a "purely dynamical metric theory of gravity", meaning a metric theory "whose gravitational fields have their structure and evolution determined by coupled partial differential field equations" (Will, 2014, §3.1.2). Will analyzes the differences between GR, Brans-Dicke theory, and Visser's bimetric massive gravity theory in terms of whether each is a "purely dynamical metric theory" or a "prior-geometric metric theory". The PPN formalism, first presented in unified form by Nordtvedt and Will, is laid out in §3.2, and in §3.3, Will explains how the PPN formalism can be applied in "the comparison and classification of alternative metric theories of gravity". For the relation between the PPN and ppE formalisms, see Yunes & Pretorius, 2009, pp. 122003–3.

LIGO astronomy is done in part by relating properties of the signal to estimated properties of the source (see, e.g., Abbott et al., 2016). Properties of the signal including frequency, polarization, and strength can be taken to indicate properties (parameters) of the black hole system, including the mass of the component black holes, the chirp mass of the system, the orbital frequency and period of the system, the distance of the merger from the detector, the position of the merger in spacetime, and so on.

But as Yunes and Pretorius observe, the process of reasoning about system parameters using LIGO data is freighted with assumptions. The template banks that LIGO uses involve techniques including post-Newtonian approximation, the quadrupole formula, and numerical relativity that bake GR into the predicted waveform (see, e.g., Capano, Harry, Privitera, and Buonanno (2016)). Even though there is no direct solution of the linearized Einstein field equations in the strong field regime, those equations are used as formal constraints on many of the templates that are used to confirm parameter estimations in LIGO astronomy. If the theoretical assumptions from GR are incorrect as descriptions of the detected system, then the signal will still be *detected*, but the conclusions drawn from it will be inaccurate. For instance, a signal might be detected but parsed as a weak signal using the current templates, and as a result the system might be inferred to be distant from the detector. But if the signal is in fact a strong signal *that is inconsistent with GR*, it may be much closer. Yunes and Pretorius observe that such a case is a real possibility given the assumptions made in template-based searches.

It is not trivial to remove the assumptions from the templates. As explained above, LIGO data consists of measurements of differential lengths of interferometer arms, indexed to time. The expected noise must be modeled and subtracted from the signal. In Hanford, Washington, where one detector is located, noise ranging from seismic vibrations due to logging trucks to shot noise from quantum effects may complicate the signal. Once a maximally 'clean' signal is obtained, a search is performed using templates covering the expected signal space. One search looks simply for a signal 'burst' picked up by multiple detectors, while another template-based search looks for a waveform with certain expected parameters. In the case of the landmark first detection, GW150914, "Two types of searches found this event: an unmodeled search designed to look for coherent "bursts" of power in both LIGO detectors and a modeled search designed to search for gravitational waves (GWs) from coalescing binary neutron stars (BNS), neutron-star black hole binaries (NSBH), and binary black holes (BBH)" (Capano et al. (2016), pp. 124007–1).[22] An unmodeled search can detect a signal, but does not have the sensitivity to show whether that signal is consistent with a waveform that would be predicted to emanate from a system with specific parameters. Modeled searches using template banks are preferable, then, in order to confirm that the detected waveform is a detection *of* a system with specific properties (Capano et al. (2016); Abbott et al., 2016).

The argument of Yunes and Pretorius is that the current framework for modeled searches contains structural features that effectively build the assumption that GR is correct into the modeled search and confirmation process. The more sensitive and specific the template-based search is, the more assumptions the method will embody. The very method that contributes to the significance of the results of the search - what makes it a five sigma result - may be masking information about gravitational waves and their sources that is inconsistent with the background theory of general relativity.

### 2.3. The parametrized post-Einsteinian framework

The parametrized post-Einsteinian (ppE) framework was devised to resolve this problem. It is intended to be close enough to GR to be able to analyze the data properly, but at the same time to be able to "measure *generic* deviations from GR predictions" (Yunes and Pretorius (2009), pp. 122003–3). Note the similarity between this terminology and Curiel's distinction between "abstract" equations, "generic" physical systems, and "concrete" and "individual" models (Curiel, 2020, §7). Yunes and Pretorius identify a class of physical systems determinable as "modifications to GR" in the late merger stages. They begin by analyzing which claims of GR can be considered valid constraints on theories in the neighborhood, and identifying those structural elements of the theory where "generic deviations" from it may be detectable. They reason,

> starting with GR binary BH merger waveforms seems sound. We are then faced with the question of how to modify these waveforms in a sensible manner. In theory, there are uncountably many conceivable modifications to GR that only manifest in the late stages of the merger. To make this question manageable, we shall guide our search for ppE expansions by looking to alternative theories that satisfy as many of the following criteria as possible: (i) *Metric theories of gravity*: theories where gravity is a manifestation of curved spacetime, described via a metric tensor, and which satisfies the weak equivalence principle. (ii) *Weak – field consistency*: theories that reduce to GR sufficiently when gravitational fields are weak and velocities are small, i.e. to pass all precision, experimental, and observational tests. (iii) *Strong – field inconsistency*: theories that modify GR in the dynamical strong field by a sufficient amount to observably affect binary merger waveforms (Yunes and Pretorius (2009), pp. 122003–2).

The idea is to build a framework within GR that is able to detect deviations from GR's predictions in the strong-field regime. The framework is complex and involves a number of extensions and modifications to the methods for analysis of LIGO data.

Perhaps the most significant observation for our purposes is that the ppE framework was designed for post-detection data analysis, not for detection or search.[23] The framework is intended as a network of relations within the theory that nonetheless allow for information that is in conflict with the theory to be observed.

The term "parametrized" does not mean, exactly, that the framework has to do with parameters of the source system (mass, spin, and so on). The term should be understood, instead, in the sense of "parametrized" models in statistics, in which substantive variables are replaced with parameters so that the range of the parameters become the focus of the analysis. Some ppE parameters are physical parameters (the chirp mass), but some are intended to measure specific assumptions of the model. For instance, in the merger and ringdown phase, Yunes and Pretorius introduce constant parameters that are "set by continuity" - that is, their predicted value is set by the assumption that the waveform will be continuous in these phases (Yunes and Pretorius (2009), pp. 122003—4). This move is akin to what Curiel describes as moving from the abstract equations to a family or "species" of physical systems. Other parameters are set by the assumption that, for instance, the motion of a system element is well described by a known analytic function. The point of "setting" parameters in this way is to generate predicted, hypothetical waveforms. These hypothetical waveforms can then be used in post-data analysis to detect deviations from theoretical assumptions and predictions - even fundamental errors about the phenomena being investigated.

---

[22] "The primary difference between these searches is that the modeled search uses a bank of template waveforms of expected signals to match filter the data, obtaining a signal-to-noise ratio (SNR) for candidate events ... Key to maximizing the sensitivity of the modeled search is that the parameters of the template waveforms are sufficiently close to the sources' parameters such that the morphology of waveforms matches that of signals. Any mismatch between signal and template leads to a loss in SNR, and down-weighting by the signal-based vetoes. Some source parameters, such as the coalescence time and phase, can be analytically maximized over, resulting in essentially no SNR loss. The remaining parameters, however, are traditionally covered by some gridding of the parameter space, in which a small but nonzero amount of SNR is lost to signals from systems not lying exactly on the grid" (Capano et al. (2016), pp. 124007–1).

[23] "We do not propose here to employ the ppE templates for direct detection, but rather for post-detection analysis. Following the detection of a GW by a pure GR template, that segment of data could then be reanalyzed with the ppE templates" (Yunes & Pretorius, 2009, pp. 122003–5).

These parameters of the ppE framework could be called "descriptive parameters", to distinguish them from the physical parameters discussed in the section above. The reason to use descriptive parametrization of the GR equations, as Yunes and Pretorius do, is to obtain control of the counterfactuals: for instance, "If the merger is continuous, then the waveform will have the following form" (given certain other assumptions). Those counterfactuals are crucial to *testing* for deviations from a theory's predictions, which is a different process from *confirming* those predictions.

The parametrization of the equations used for detection and simulation of waveforms is key to the ppE framework. GR makes predictions about the values - and the ranges of values - for certain variables. The parametrized equations replace the variables with parameters, which puts the focus on questions like the dependence of parameters on others within the system, and the structure of the dynamics of the system. In some cases, the dynamics is reduced almost entirely to kinematic structure and then the dynamics of GR is re-introduced as a special case.[24]

Yunes and Pretorius provide an anatomy of a binary black hole coalescence, including GR based models of the inspiral, merger, and ringdown phases of the coalescence. Following this, they introduce modifications to the fundamental equations used in LIGO detections and template building. One might ask why Yunes and Pretorius would begin with models in GR, since they are proposing a kind of formal language that is broader than GR. The reason they do so, I believe, is that their presentation focuses on how the parts of the GR formalism are related to each other. Parametrized models allow for relationships between formal elements of the theory to be made more apparent, and Yunes and Pretorius use this to good effect in modeling the conservative and dissipative dynamics of black hole mergers. The first GR model they consider consists of "modifications of the Fourier transform of the response function during the inspiral arising either from changes to the conserved binary binding energy (the Hamiltonian …), or changes to the energy balance law. The former are essentially modifications to the *conservative* dynamics, while the latter modifies the *dissipative* dynamics" (Yunes & Pretorius, 2009, pp. 122003—8).

In this section (IIIA), Yunes and Pretorius focus on how the conservative and dissipative dynamics are derived from the binary Hamiltonian and from Kepler's law, in the former case, and from the balance law and the perturbation of the stress-energy tensor, in the second case. They go on to perform similar modeling tasks for the merger and ringdown phases. In each case, they find a parametrization that is consistent with the GR modifications to classical mechanics. But they do more than this:

1. In several cases, they allow for the replacement of a specific GR formalism with an analytical expression in terms of a known function, which allows for a "generic" statement of the equation that's in play (e.g., Yunes & Pretorius, 2009, pp. 122003—8).
2. In the case of their modifications to the conservative dynamics in the inspiral phase, they provide a generic parametrization that encodes only the predictions from the Hamiltonian and Kepler's law - but then they allow for the GR and post-Newtonian predictions to be recovered within that formalism, by constraining the values of the parameters, for instance (Yunes & Pretorius, 2009, pp. 122003—9).

The theoretical language and substantive predictions of GR are being replaced with more general formalisms that appeal to known functions that can be inverted, that replace variables with parameters, and, in general, that replace the GR strong-field equations and structure with a broader scaffolding for rigorous post-data testing. Once key equations are replaced with their parametrized forms, the ppE framework can be put together as a whole. The main virtue of using this framework is that, if it is used for post-detection data analysis, ideally it would catch the cases of interest to Yunes and Pretorius: the cases in which a detection is made but the data contain information inconsistent with GR. Patel (2019), for instance, provides a recent set of computational methods for post-data analysis using the ppE formalism.

Hempel is right that, if we switch between theories in physics by interchanging the observables and leaving only uninterpreted theoretical terms and symbols, we may not preserve the empirical knowledge and tacit knowledge encoded in those physical theories. But Carnap, Hilbert, and Curiel are also right that terms with a specific role or place in a structure - like Hooke's constant or the terms 'line' and 'plane' in geometry - can be made abstract by turning the theory itself into a "generic structure" with multiple possible manifestations. The ppE method uses parameters to construct a class of dynamical systems of just this kind. That methods defines a class of structures, and testing can be set up to try to find experimental ways to decide between them.[25]

A closing note: Einstein likely would have learned of all this with pleasure. Einstein's frequently expressed wish was for the theory of relativity to be as open to testing as possible.

### 3. Confirmation versus testing in general relativity

The summary above of recent progress in gravitational wave astronomy shows that, depending on the methods used, the data from advanced LIGO research can be taken to confirm, or to test, general relativity - but the methods must be constructed carefully. The ppE framework constructed by Yunes and Pretorius provides a specific example of constructing a broader framework for post-data testing, a broader framework capable of detecting flaws in the structural assumptions of the theory with respect to the observed data.

While many would reject a strict distinction between theoretical and observational vocabularies in the sense of the early Carnap, Stein's work points to the fact that an in-practice distinction between *confirmation* of a theory and *testing* of that theory via

---

[24] See Curiel (draft) for discussion of the relationship between kinematics and dynamics in the structure of spacetime theories.

[25] Here, I should note the obvious precedent of the semantic tradition in philosophy of science, exemplified by Stegmüller, Sneed, Suppes, and perhaps Hesse, and with precursors including Hertz and Hilbert.

observation is necessary. As Carnap himself remarks in "Testability and Meaning",

> a sentence may be confirmable without being testable; e.g. if we know that our observation of such and such a course of events would confirm the sentence, and such and such a different course would confirm its negation without knowing how to set up either this or that observation (Carnap, 1936, pp. 420–421).

General relativity has been very well confirmed, in the weak and now in the strong field regimes. Testing is a very different matter. In the case of advanced LIGO, the need for analysis of the theory before further testing is great. The "observations" of gravitational wave astronomy require so much theoretical structure to formulate that the theory must be re-formulated for tests to detect, not only deviations from the theory's predictions, but information that observations made using the theory will obscure.

It is not possible to provide a universal theoretical language to relate GR to LIGO observations. Thus, there is no ready-made *framework* for testing - which is entirely distinct from whether there is a way to test whether a given set of observations can be warranted to be a confirmation of GR using the likelihood function. Just because there is no ready-made framework, however, does not mean that there cannot be a framework that is constructed within the theory itself. The joint testing of the classical and new quantum theories is one example of a relation between theoretical languages for the purposes of theory testing. Nordtvedt and Will, and then Yunes and Pretorius, construct alternatives - the PPN and ppE frameworks - where GR itself is turned into a framework for testing through parametrization.

The ppE framework is the first step toward this scaffolding for testing, a scaffolding constructed within the theory itself, but using parametrization and other formal methods to make the theory more 'generic'. Yunes and Pretorius re-formulate the models of black hole coalescences using a "parametrized" version of the GR equations, which allows for a descriptive parametrization, which in turn allows for counterfactual reasoning about what the waveforms would look like given different formal assumptions. As a result, using different values for the parameters and incorporating different constraints, one can construct a more sensitive instrument for the testing of the theory, out of the theory itself, which becomes a means of determining a class or family of related theories or structures.

Even before the hypotheses of the theory are tested, Yunes and Pretorius suggest that the theory itself must be investigated to see whether its basic assumptions about the phenomena, influenced by a background theory, are descriptively inaccurate of the data under investigation. The question is whether the formal assumptions of a theory are descriptively, formally, or statistically adequate to the data under investigation. If a researcher encodes strong assumptions into the entire modeling framework, the problem is not that a specific test of a hypothesis will be flawed, it is that the entire framework used to formulate hypotheses may be misleading - and that we do not have the tools to detect that it's wrong until the framework is modified.[26]

Yunes and Pretorius construct a set of related theories determined by distinct assumptions and by the dynamical systems they embody. For instance, in the case of the inspiral, the conservative energy dynamics are modeled using the Hamiltonian and Kepler's law as the fundamental relations, and fewer assumptions are made about the dynamics of the system. The parametrized models Yunes and Pretorius construct are not general relativity. But they are not *not* general relativity - they are not derived from a rival theory. Instead, key structural elements of GR are made, at the same time, more general and more inferentially rich, in the sense that the "generic" versions of the equations allow for multiple interpretations, derived hypotheses, and empirical possibilities. In Yunes et al. (2016), the ppE framework is used to support reasoning about *rivals* to general relativity, and extensions of current theorizing about gravitational waves to future detections:

> Events GW150914 and GW151226 [the first two LIGO detections] are fantastic probes of theoretical physics that have important implications for certain aspects of extreme gravity, but unfortunately not all. Future detections of GWs from NS [neutron star] binaries will allow us to probe different aspects of extreme gravity. The prime examples of this are theoretical models where gravity is described by a metric tensor with evolution equations that differ from the Einstein equations only through a modified "right-hand side" that depends on the matter stress-energy tensor. Examples of such theoretical models are Eddington-inspired Born-Infeld gravity and Palatini f(R) theories… Other examples include the activation of certain scalar or pseudoscalar fields in the strong-gravity regime sourced by dense matter, such as Brans-Dicke theory, the scalar-tensor theory extension of Damour and Esposito-Farèse, and f(R) models as they can be mapped to scalar-tensor theories (Yunes et al., 2016, pp. 084002–29 - 084002–30).

To test the theories mentioned, first it is necessary to establish exactly how the theories differ formally from Einstein's equations, and then to model what evidence for each of the theories would look like. This may consist in the first instance of formal reasoning about the internal structure of the theory, before moving on to test the hypotheses and models empirically.

Yunes and Pretorius (2009) parametrize the equations of general relativity, and the waveform models derived from them, to remove bias. In this way, Yunes and Pretorius make the theory more broadly empirically testable, not less. But they do so by modifying the formal laws, principles, definitions, and models of the theory. By parametrizing the theory elements, Yunes and Pretorius make it possible to test the theory more rigorously using experimental observation, and to draw more secure inferences about the extent to which the theory has been tested. The potential of the ppE framework remains untapped to a great degree, but is in use in recent expositions of GR and the LIGO project.

The construction of the ppE framework does not involve any new experimental technique, and yet it extends the experimental reach of the theory. It does so precisely by making it possible to "get the laboratory within the theory": to engage in a priori reasoning about the empirical reach of the theory, trying to pin down the breaking points of theories, the regimes in which they are no longer applicable, and the species (in Curiel's sense) of systems in which they may break down. Moreover, constructing theories in which the 'abstract' and 'generic' equations and variables of the theory are clearly, explicitly identified, allows for a deeper understanding of the structural relationships of that theory: which in turn allows researchers to pinpoint even more interesting problems.

The testing method provided via the ppE framework is an excellent reflection of Stein's argument that moving to Carnap's 'first volume' can be a more empirically well founded move in terms of providing a broader base for testing than gathering more and more specific observations. If the modeling and detection processes in gravitational wave astrophysics encode a framework of fundamental assumptions that are incorrect, then, paradoxically, the more data is gathered, the farther from the correct analysis we will be. Only by re-conceiving the fundamental relationship between the theoretical and observational vocabularies of the theory can our observations be a true test of the theory, and indicate avenues for future research.

---

[26] See, e.g., the statistical model specification approach found in Spanos, 2011. Spanos & McGuirk, 2001 traces statistical model specification to the work of R.A. Fisher (pp. 1168–9).

### CRediT authorship contribution statement

**Lydia Patton:** Writing - original draft, Writing - review & editing.

### Acknowledgements

I would like to thank, and congratulate, Erik Curiel, James Weatherall, and Thomas Pashby for organizing this special issue in honor of Howard Stein, and to thank them and James Ladyman for editorial guidance. Two anonymous reviewers worked with me, really, to revise the paper, and I am very grateful for their engagement with the project and for their insights into how to improve the project. Don Howard suggested that I work on the LIGO results, and I am very glad he did. Aris Spanos has provided incredible support for my research into data analysis and the LIGO project, including working through LIGO papers and analysis of LIGO data with me. He and Deborah Mayo have been frequent and fascinating interlocutors about the questions of data analysis and testing that underlie this project. Jamee Elder and I worked through several of the papers cited here together, and her insights into the papers were astute and valuable. Versions of the paper were presented at the inaugural conference of the Black Hole Initiative, the Cambridge University Philosophy of Science seminar, the American Mathematical Society, and the Munich Center for Mathematical Philosophy. The discussions there pushed toward a much better definition and understanding of the overall structure and argument of the paper, and I am very grateful for the discussions there. In this context, it is an especial pleasure to recognize the contributions of Hasok Chang, Richard Staley, Andrew Strominger, Ramesh Narayan, Peter Galison, Erik Curiel, J. Brian Pitts, Liba Taub, Marta Halina, Neil Dewar, Jacob Stegenga, Alexander Reutlinger, and Patricia Palacios. I am sure I am forgetting some people; my apologies. I owe a great debt to the LIGO Collaboration, who make a great deal of their data publicly available, and who are enthusiastic about discussing and disseminating their work. In particular, I am grateful to Jonah Kanner and to the LIGO scientists who organized and led the LIGO Open Science Workshop in Pasadena in March, 2018. Finally, but certainly not least, I would like to recognize Heta Patel and Eric Wuerfel for their fantastic work on our project on LIGO that stretched from 2017 to 2019, and Benjamin Jantzen for facilitating our work on this project.

### Appendix A. Supplementary data

Supplementary data to this article can be found online at https://doi.org/10.1016/j.shpsb.2020.01.001.